# Low Order Adaptive Optics with Very Faint Reference Stars.

Craig Mackay [1]

[1] Institute of Astronomy, University of Cambridge, CB3 0HA, UK
email: cdm <at> ast.cam.ac.uk

**Abstract.** It is widely believed that adaptive optics only has a role in correcting turbulent wavefronts on large telescopes using very bright reference stars.
Unfortunately these are very scarce and many astronomical targets require wavefront correction to work over much of the sky. We therefore need to be able to use very much fainter reference objects. Laser guide stars in principle can allow 0.1 arcsecond resolution but have a number of severe technical problems that limit their application. Our aims are to provide imaging at even higher resolution than Hubble.
Lucky Imaging completely eliminates the tip-tilt errors in astronomical wavefront detection. Most of the power that remains is in low order, large scale structures. These may be detected with high sensitivity using photon-counting EMCCD detectors working at high frame rate, up to ~100Hz. With a new design of curvature wavefront sensor, wavefront errors may be measured and corrected to give near diffraction-limited performance on large ground-based telescopes in the visible. Reference stars (and reference compact galaxies) fainter than I~17.5 mag may be used routinely. This paper will describe how these work, what detector and other hardware is needed and what software should be used to measure the wavefront errors and drive deformable mirror hardware. The software techniques that are used are those routinely applied for MRI and CT imaging. They are fast and relatively easy to implement. The net effect is that imaging systems can be constructed that improve substantially over Hubble resolution from the ground for a relatively modest sum of money.

**Keywords:** adaptive optics, Lucky Imaging, curvature wavefront sensor, Shack-Hartmann sensors, radon transforms.

## 1    Introduction

The Hubble Space Telescope (HST) delivered routinely for astronomers images of extraordinary quality. Ever since the invention of the telescope the angular resolution achievable from a ground-based visible or near infrared telescope has been limited by atmospheric turbulence. On a good site that is typically ~ 1 arcsec, sometimes better.

It is now more than 25 years since the HST was launched and inevitably it cannot continue to operate fully forever. The refurbishing missions for the telescope are no longer possible with the retirement of the US Space Shuttle programme. Increasingly astronomers are looking to the future and how images as good as the HST might be available and indeed how even sharper images might be obtained from ground-based telescopes.

The James Webb Space Telescope is currently scheduled to be launched in the second quarter of 2019. As a telescope it is capable in principle of higher resolution in principle. It has 18 hexagonal mirror segments for a combined mirror size of about 6.5 m diameter. This can be compared with the HST which has a 2.4 m diameter single mirror. However the JWST is primarily intended for near infrared observations. In most applications it will give images with poorer resolution because the pixels used in the detectors are coarser than those of the HST.

For many years astronomers have tried to deliver improved images on ground-based telescopes. Most of these methods use techniques broadly called Adaptive Optics. The resolution of any telescope is set by the variance in the phase over the entrance pupil of the telescope. Noll (1976) showed that the variance depends on the ratio of the telescope diameter to the atmospheric cell size, $r_0$, where $r_0$ is the Fried seeing parameter (Fried, D.L., 1978, JOSA, 68,1651). The turbulent spectrum is assumed to follow the Kolmogorov law or at least variations that are fairly close to it. The phase characteristics across an instantaneous pupil measurement may be expressed in terms of the sum of Zernike polynomials or the very similar Karhunen-Loeve modes. The paper by Noll (1976) shows that for a typical turbulent spectrum the power in each Zernike polynomial rapidly reduces as the polynomial mode gets bigger. This is shown in the following table 1.

**Table 1:** the Zernike power spectrum of atmospheric turbulence (after Noll, 1976).

| Zernike Terms Removed | Power in Each Zernike Term | Residual Power Fraction |
|---|---|---|
| 1 | 0.0 | 1.0 |
| 2 | 0.435 | 0.565 |
| 3 | 0.435 | 0.130 |
| 4 | 0.022 | 0.108 |
| 5 | 0.022 | 0.085 |
| 6 | 0.006 | 0.063 |
| 7 | 0.006 | 0.057 |
| 8 | 0.006 | 0.051 |
| 9 | 0.006 | 0.045 |
| 10 | 0.006 | 0.039 |
| 11 | 0.0025 | 0.037 |
| 12 | 0.0025 | 0.034 |

If we are able to take images fast enough and there is an adequately bright reference object in the field of view then each frame may be registered before being added to the existing sum of the images. In that way a very large part of the turbulence is removed (the tip-tilt component) reducing the variance across the pupil to less than one seventh of its original value. That in turn effectively increases the size of the characteristics seeing scale, $r_0$. This improves the effective seeing by a factor of around 2.7 relative to that normally achieved under the same conditions. Removing the tip tilt in this way is the basis of Lucky Imaging. The results of many observing runs (see, for example, Baldwin et al., 2008) has confirmed substantially the predictions of Fried and Noll. At the time of writing this there were over 350 papers already published with the phrase "lucky imaging" in their abstract.

The quest to be able to measure and then remove as much of the turbulence as possible has led to the development of different ways of sensing the wavefront. Once we have measured the deviations from flatness in the phase of the wavefront we can use a deformable mirror to correct for the turbulence in the atmosphere. This is the principle that underpins most Adaptive Optics instruments.

The high quantum efficiency of silicon digital detectors such as CCDs has made them the detector of choice for many years for adaptive optics detector systems. For many years good readout noise, essential for good sensitivity with faint reference objects, could only be achieved with slow-scan CCDs. It therefore became very popular to operate these with considerable amounts of binning in both directions to give an acceptable frame rate. By putting a relatively small area CCD behind an array of small lenses it was possible to image the light from a series of smaller apertures across the pupil. These apertures produced an image from each lenslet. By measuring the displacements of the images from the rectangular grid corresponding to the lenslets centres it was possible to work out the distortion across the pupil and hence to use the net wavefront curvature across the pupil that had to be corrected for. This is the basis of the Shack-Hartman wavefront sensor.

The Shack-Hartmann wavefront sensor has been used for many years. However it does have fundamental limitations which make it relatively unsuitable for work on faint reference objects. In practice the number of lenslets used in a Shack-Hartmann sensor may be relatively large, often many hundreds in total. This means that the light from the reference star is split up amongst several hundred detector areas. If the star is fairly bright enough to be detected reliably with one lenslets it is then the images from every lenslet may be detected simultaneously. The Shack-Hartmann sensor goes from being an effective and working device at one brightness level whereas slightly fainter stops at working entirely.

In many cases the astronomer would be entirely happy with a less precisely measured set of wavefront errors. Unfortunately the Shack-Hartmann sensor used on a slightly fainter target would produce no output. The sensitivity of the Shack-Hartmann sensor is therefore very poor when compared to the light gathering power of the entire telescope. The need to use very bright reference stars in order to get a good wavefront estimation restricts the use of Shack-Hartmann sensors to a very small fraction of the sky. With stars that are perhaps in the range of 11-13 magnitude the fraction of the sky that is accessible is extremely small indeed, much less than 1%.

AO system engineers have also worked on the use of laser guide stars. Here the idea is that by using a powerful ground-based laser aimed close to the science target of the telescope it is possible to make a compact bright area visible to the telescope with the correct filtering. These artificial laser guide stars may be used in place of a natural reference star should none be available. Laser guide stars break into two separate types. Those that use the sodium layer at an altitude of around 90 km and Rayleigh Beacons which rely on the scattering of light by molecules in the lower atmosphere.

Sodium Beacons are able to provide better wavefront references because the sodium layer is so much higher than that used for Rayleigh Beacons. The sodium arises as a consequence of meteorite showers entering the atmosphere. As the meteorites burn up they contribute small amounts of sodium ions into the upper atmosphere which recombine. These sodium atoms may be stimulated into emission by an appropriate wavelength of laser light. Unfortunately variations in the amount of sodium injected into the mesosphere make the performance of sodium lasers relatively unpredictable. They depend very much on the availability of sodium from season to season. Laser guide stars are also difficult to use should there be any opacity in the atmosphere (such as thin high altitude cloud, or haze or dust). Despite the considerable effort put into these laser guide stars they have not yet been able to demonstrate ground-based telescope resolution equalling that of the HST routinely.

Recently the European Southern Observatory's VLT has been fitted with a quad laser system. This produces a mini constellation of laser guide stars around the target area. For laser guide stars to work even in theory with a telescope as big as the European Large Telescope multiple guide stars are necessary.

Despite the eye watering amount of money spent around the world on AO systems and laser guide stars , results have been distinctly mixed. The more successful ones have improved resolution over small fields of view by a moderate amount on modest size telescopes. However although the amount of money spent on these systems is now comparable to the HST total cost to date it is difficult to feel that they are even close to being competitive with the HST.

This paper is directed towards developing the technology for an entirely different way to measure the wavefront errors across the pupil. It attempts to measure the curvature in the wavefront directly and to use that to drive a deformable mirror in much the same way as is done with the Shack-Hartman systems.

## 2    The Curvature Wavefront Sensor.

The basic concept of the curvature wavefront sensor is not new. The difficulty, for many years, has been in finding detectors that were able to make best use of that instrumental set up. Our work has been directed towards removing a large part of the atmospheric phase errors across the pupil of the telescope and then relying on Lucky Imaging techniques to provide diffraction limited images by using a fraction of the total images obtained. This combination of low order AO plus lucky imaging has already produced the highest resolution picture ever obtained in the visible or infrared from anywhere, either ground or space based. Our goal is to repeat that success but to allow it to be achieved with much fainter reference stars.

The original test of the combination of low order AO plus Lucky Imaging was carried out on the Palomar 5 m telescope using their PALMAO system (Dekany, R. G., et al, 2000). The PALMAO system is now rather old and has recently been replaced. It uses a Shack-Hartmann wavefront sensor and a Xinetics deformable mirror. However for our purposes it provided low order adaptive optics reliably given a relatively bright reference star. We observed in the M13 globular cluster using one of the bright stars in the middle of the cluster for reference. Mounting our Lucky Imaging camera behind PALMAO we obtained images shown in the figure. The observing conditions were good with seeing of about 0.65 arcsec. The pair of images in the figure shows firstly the image that would have been obtained simply by summing the consecutive sequence of images. The other image was obtained by lucky imaging, selecting 20% of the images. The resolution in that image is about 35 milliarcseconds in I band with a peak Strehl ratio of about 17%. Although taken over 10 years ago this image is still the highest resolution image ever taken in the visible or IR.

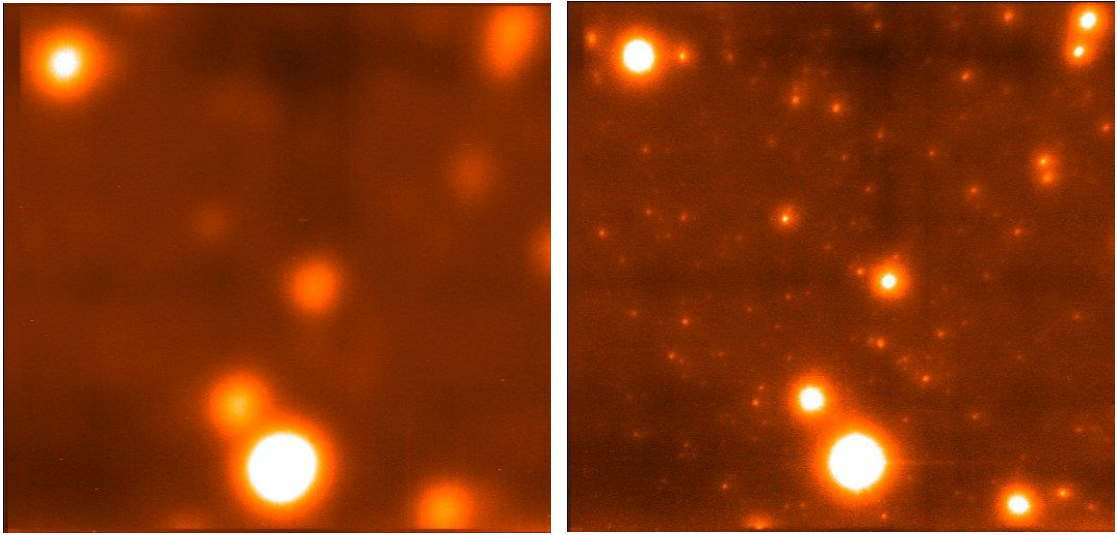

**Figure 1.** Images taken on the Palomar 5 m telescope with the PALMAO low order adaptive optics system combined with a Lucky Imaging camera. The images were taken of the globular cluster M13 in I-band. The seeing was about 0.65 arcseconds. The left-hand image is simply the sum of the unprocessed images under normal natural seeing. The right hand image is a 20% selection of the images recorded after correction by the adaptive optic system. The resolution of this image is ~ 35 milliarcseconds making it the highest resolution image ever recorded of faint targets anywhere, either from space or from the ground.

If we think about the way that light propagates through the aperture of the telescope then considering only that light we see that the telescope pupil is uniformly illuminated itself but if we image the appearance of that light as we gradually increase the distance from the pupil plane we find that uniform appearance of the pupil breaks up into speckle-like structures. These are not speckles in the conventional use of that word as the elements into which the seeing profile is broken. We can also see from the figure that as the distance from the pupil plane is increased the characteristic scale of these fluctuations becomes larger.

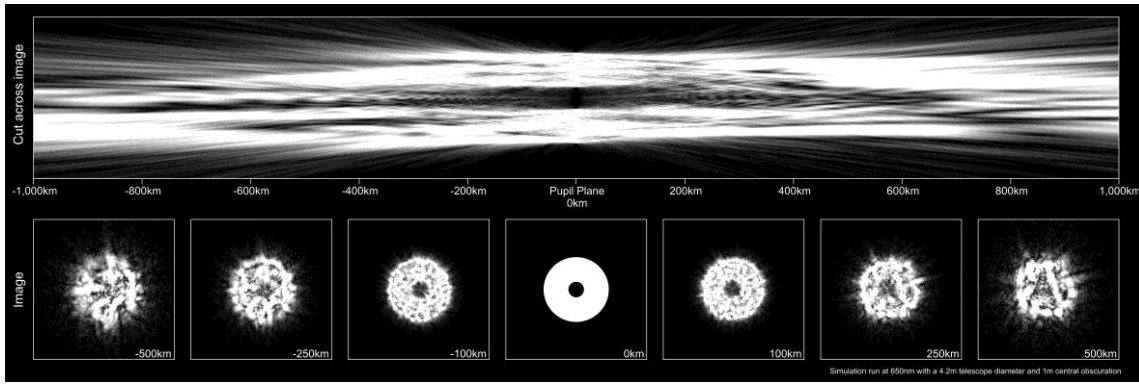

**Figure 2.** This shows the way that light that passes through the pupil of the telescope changes its appearance at different distances away from the pupil. The light intensity is uniform across the pupil although the phase is not uniform due to atmospheric turbulence and other effects. As the imaging planes move further from the pupil the image breaks up into speckle -like structures. The further away from the pupil we image, the structure of the images become increasingly dominated by large-scale features. (Guyon et al, 2008, 2010).

We can understand how a curvature sensor works if we think of the images that would be obtained by detecting the light on either side of the pupil plane (see figure 1). If we have one patch on that defocussed pupil plane which is bright and the corresponding patch on the other side of the pupil is dark then we can tell that the wavefront was diverging at that point. Similarly a dark patch that becomes brighter after propagating through the pupil must be a converging element of the wavefront. By comparing the images on either side of the pupil plane we can estimate the curvature across the entire pupil. The distance at which it is easiest to estimate the curvature depends on the scale of turbulence that we wish to remove. Our interest is particularly in the use of faint reference objects. Our requirements then are very much on larger, coarser scale on these extra focal images.

There are obvious advantages in detecting the images on either side of the pupil simultaneously. The extra focal images will change relatively rapidly as the turbulence across the pupil changes and any delay between imaging on alternative sides could cause difficulties in determining a reliable wavefront curvature estimate. In practice we do this by splitting the light that has passed through the pupil and reimaging it on the bench of the AO instrument. The images are detected with a photon counting electron multiplying CCD (EMCCD, manufactured by E2V-Teledyne, Chelmsford, UK). These devices are very similar to conventional CCDs but have an additional high gain multiplying register which amplifies the signal without increasing the electronic readout noise of the device. By running these detectors with a high enough gain to deliver a signal-to-noise of ~10 for an average individual photon they may be distinguished unambiguously from any background.

## 3    Curvature Wavefront Detector Reconstruction Methods.

The methods that may be used to derive wavefront curvature from extra focal pupil images have been discussed at length by van Dam et al. (2002) this paper is worth detailed study. It views the intensity of the propagated wavefront as essentially a probability density function (PDF) for the probability of photon arrival. The wavefront aberrations distort the PDF of photon arrival. By inverting these distortions estimates of the aberration in the wavefront may be extracted. The method is easier to follow in one dimension but in two dimensions it works also very easily but does require the use of Radon transforms to unscramble the wavefront distortions.

Radon transforms are used very much in tomographic applications such as x-ray or MRI scanning systems. A reasonably straightforward account of radon transforms may be found in the documentation for Matlab. In tomographic systems the scanned sample is viewed as a projection and by analysing projections from different angles it is possible to infer what the original distortions were. Marcos van Dam created a suite of Matlab code to allow this to be tested. We have extended this in order to check a wider range of validities, but the essential code is his.

We have carried out extensive simulations based on this code. A turbulent screen is established using the methods developed by Rachel Johnson. Johnson prefers not to use Fourier methods for this but rather the Midpoint Displacement Algorithm developed by Lane et al. (1992) (Waves Random Media, volume 2, page 209, 1992). The simulated wavefront is then propagated in opposite directions to a distance which is adjustable to best represents the turbulent conditions and telescope diameter. The images are processed using Radon transforms and the fitted wavefront expressed as a sum of Zernike polynomials. The code properly takes into account photon shot noise. The Zernike polynomials derived may be combined to give the recovered wavefront. That is then subtracted from the original simulated wavefront and the differences passed to a Lucky Imaging processor.

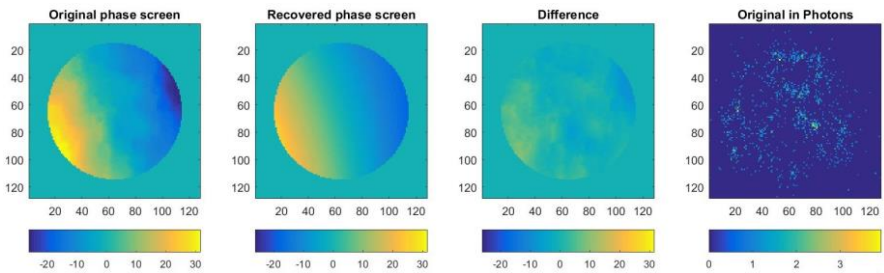

**Figure 3.** The procedures used for the wavefront fitting are shown here. The original phase screen is shown in the left-hand image and its amplitude is shown as the photon pattern in the right-hand screen. From the original phase screen the analysis produces a recovered phase screen which is very close to the original one on a large scale. When the tour subtracted the difference in the third frame is very much less than in the original. The correction has worked so far.

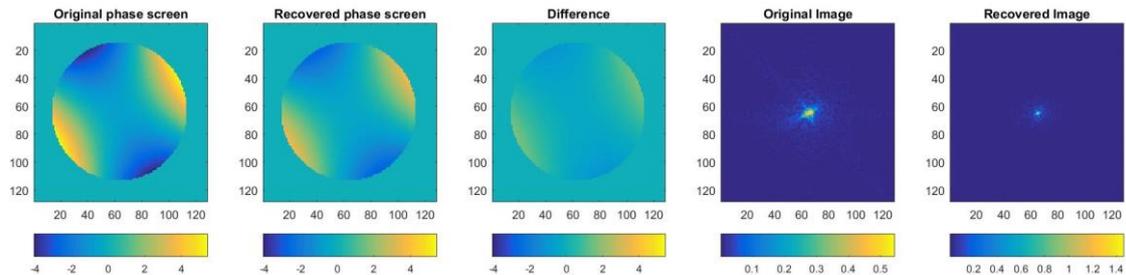

**Figure 4.** The next stage is to remove the tip tilt component from the phase screen shown in Figure 3. Again we see that the recovered phase screen looks very similar to the original phase screen and that the difference now is even smaller. That difference image gives rise to the recovered image in the fifth screen on the right. In this group the original image shown comes from the original phase screen including with tip tilt removed only.

Lucky Imaging simply carries out the image selection on the basis of phase variance across the residual phase pattern. The images are then shifted and added into different quality bins. The key to judging the success of this process for any particular set up is to compare the histograms of the variance across the corrected pupil plane before and after the processing. The differences can be quite dramatic and are shown below.

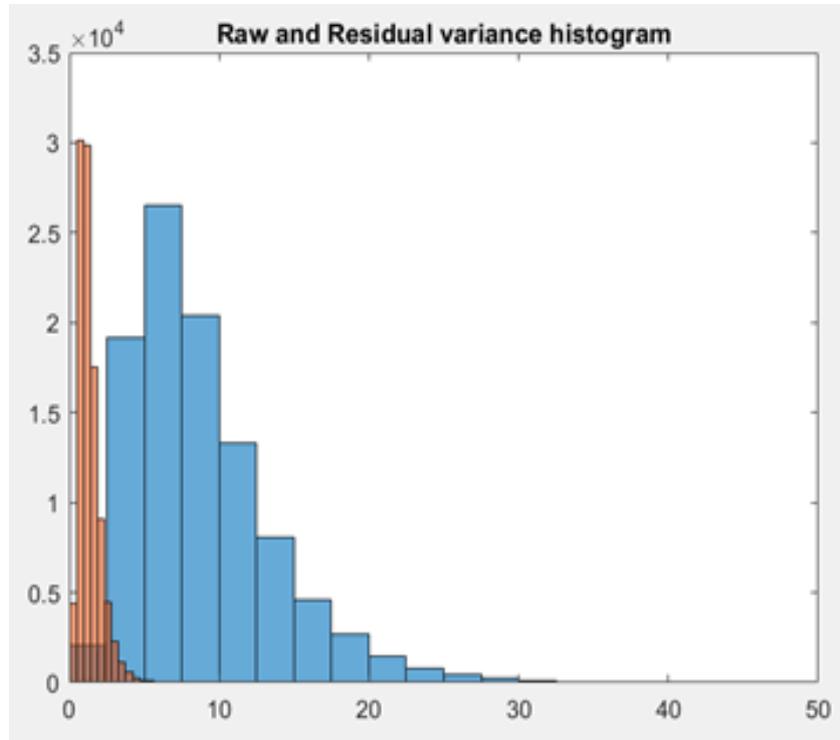

**Figure 5.** This plot shows the histogram of the variances of a large number of test wavefronts. The blue histogram shows the variances of the raw data and the brown histogram shows the variances after AO correction only.

An alternative way of inspecting the data is shown in the Figure 6:

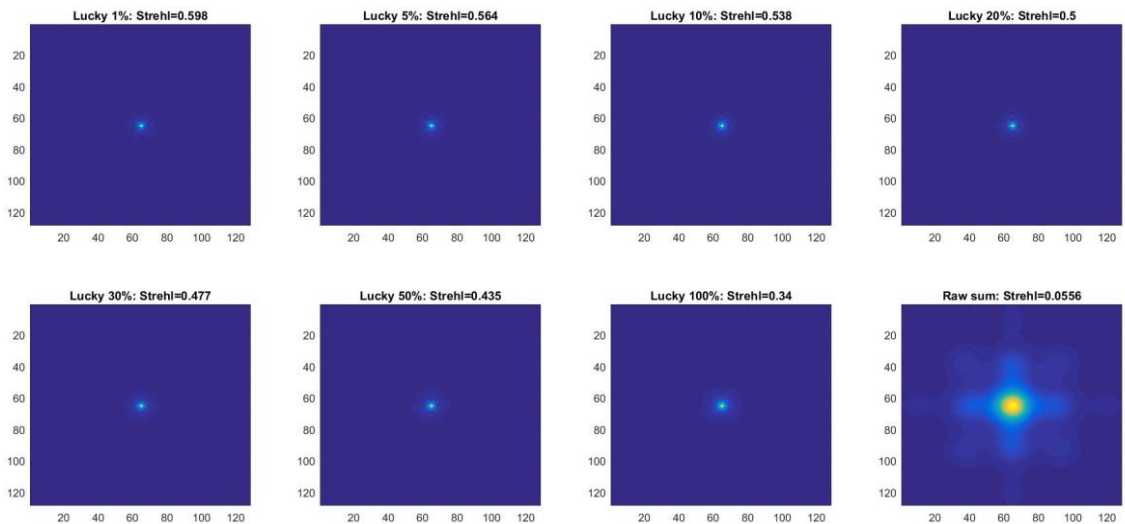

**Figure 6.** These eight images show the effects of different levels of quality selection that is an important part of the Lucky Imaging procedure. The bottom right hand image shows the raw data as generated by the wavefront simulator. Each frame is then corrected and added to each category for which it meets the required selection quality. We see that even with 100% selection the image quality is substantially improved over the original. With increasingly selective selection the image quality is further improved. The Strehl ratio obtained in each quality been is marked at the top of each box.

These simulations were carried out for a 4.2 m telescope with seeing of 0.6 arcseconds, the median seeing for the La Palma Observatory where the William Herschel Telescope is located. The images show different percentages of selection and the appearance of the images together with their effect of Strehl ratio. One-dimensional plots across these images may be taken and are shown as an example in the following image:

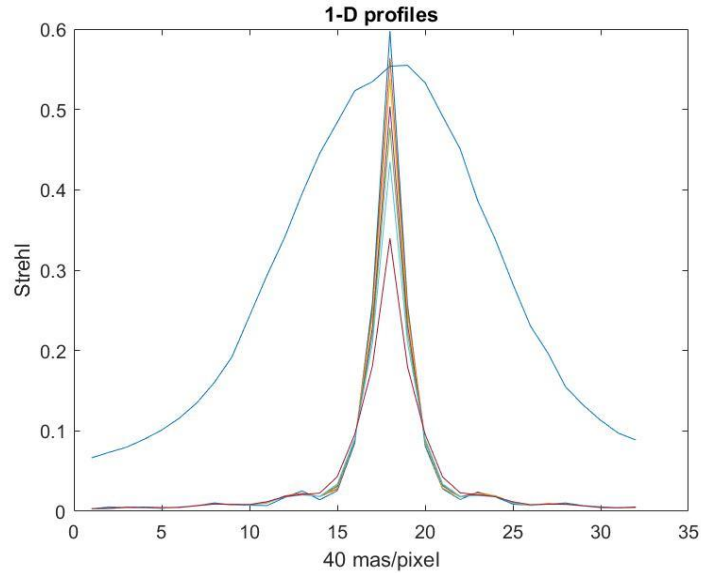

**Figure 7.** One-dimensional profiles through the eight images of Figure 6. The broad blue line is magnified x10 for clarity. The other profiles show the cuts from 100% selection (the lowest one) up to the top at 1% selection.

When trying to analyse what the net effect on image quality might be on a real telescope we need to look at the variance in each of the bins and then plot that against the number of photons per frame that were used in the simulation. Broadly speaking, a variance of less than 1 $rad^2$ gives images which are essentially diffraction limited. With a variance of less than 2.5 $rad^2$, the point spread function is about twice that of the diffraction limit at FWHM.

We found that particularly at low photon rates (consequent on using faint reference object) attempting to use too many Zernike modes to describe the wavefront gave poorer reconstruction. That was because the power in those higher modes was actually very small. Unless there were a very large number of photons per frame fitting those higher Zernike modes would be doing no more than fitting to the random position of individual photons or groups of photons in the frame. Generally that degraded the fit quality. We found that the turbulent power is almost entirely concentrated in the first six Zernike modes was critical. We have shown in table 1 that the residual turbulence for modes larger than six is barely 6% of the total and, in any individual mode, only amounts to 0.6% or six parts in 1000. Unless extremely large photon levels are present in the recorded wavefront, trying to fit such modes will simply worsen the overall quality of the fit.

It is worth noting that these techniques do not require a point source to be the reference object. When working at faint magnitudes we find that at high galactic latitudes there are generally more faint compact galaxies than there are stars. Therefore to be able to work with compact galaxies as reference objects considerably increases the number of reference objects available. The results for a 4.2 m telescope simulations suggest that with a photon rate of around 200 photons per CWFS frame, correspond-ing to reference object magnitudes of I~ 17.5-18, about 30% of the images are essentially diffraction limited (about 40 milliarcseconds FWHM) and nearly 80% give image profiles with FWHM twice that (about 80 milliarcseconds).

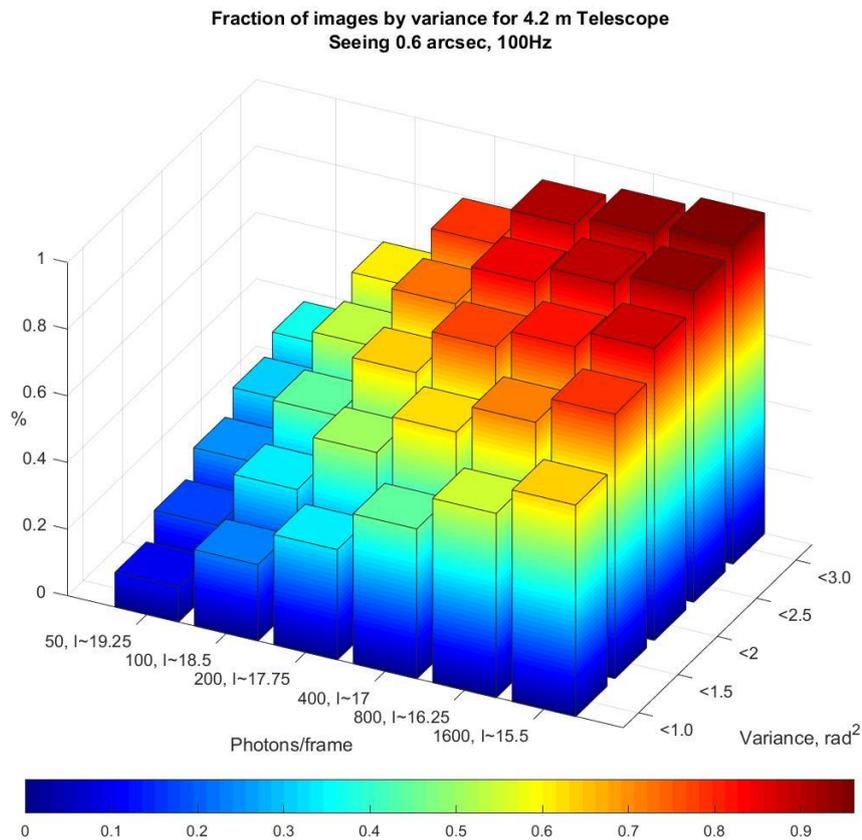

**Figure 8.** The fraction of images in different quality bins of variance compared with the photon levels per CWFS frame, together with an approximate estimate of the magnitude of star or compact galaxy that would produce such a photon rate on a 4.2 m telescope. The seeing was set to 0.6 arcseconds and 100 Hz wavefront sensor.

We can now use those same simulations and apply them to a 2.5 m telescope in a good site. Our work suggests that a 2.5 m telescope can be made essentially diffraction limited using reference stars fainter than I ~ 18, corresponding to about 60 photons per frame.
A similar analysis has been carried out and the image below shows the fraction of images by variance with a 2.5 m telescope, again with 0.6 arcseconds seeing, 100 Hz frame rate and only six Zernike modes to be fitted.

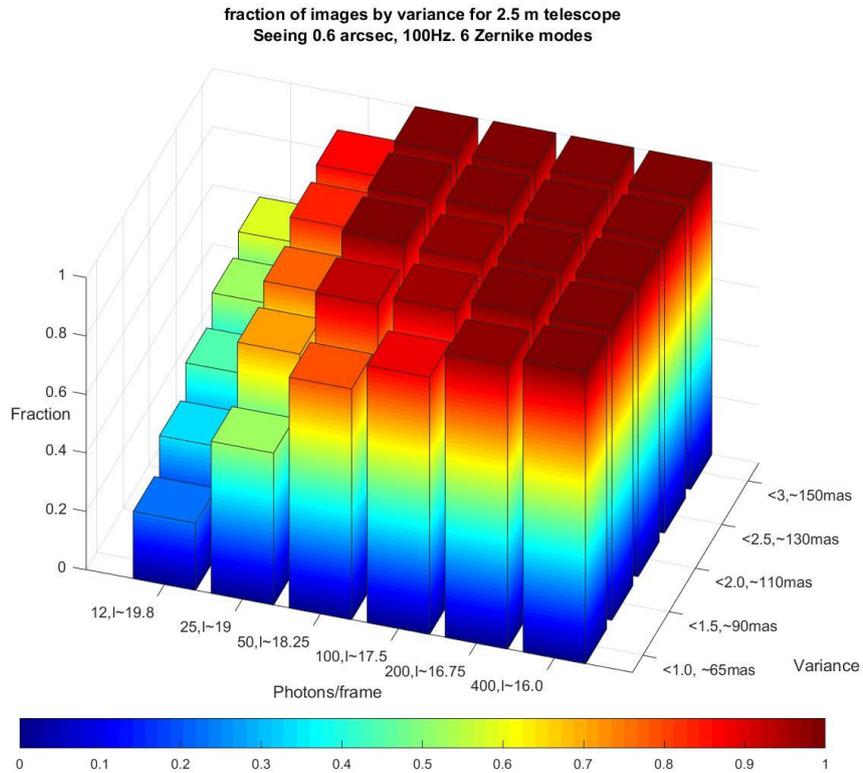

**Figure 9.** The fraction of images in different quality bins of variance compared with the photon levels per CWFS frame, together with an approximate estimate of the magnitude of star or compact galaxy that would produce such a photon rate on a 2.5 m telescope. The seeing was set to 0.6 arcseconds and 100 Hz wavefront sensor.

The techniques will work to a degree with an 8 m telescope. The photons per frame that are needed to provide satisfactory correction are significantly larger and the fraction of images that are reasonably close to the diffraction limit of 20 milliarcseconds are significantly smaller than for a 4.2 telescope. Nevertheless good correction may be obtained and a very high angular resolution achieved even in the visible.

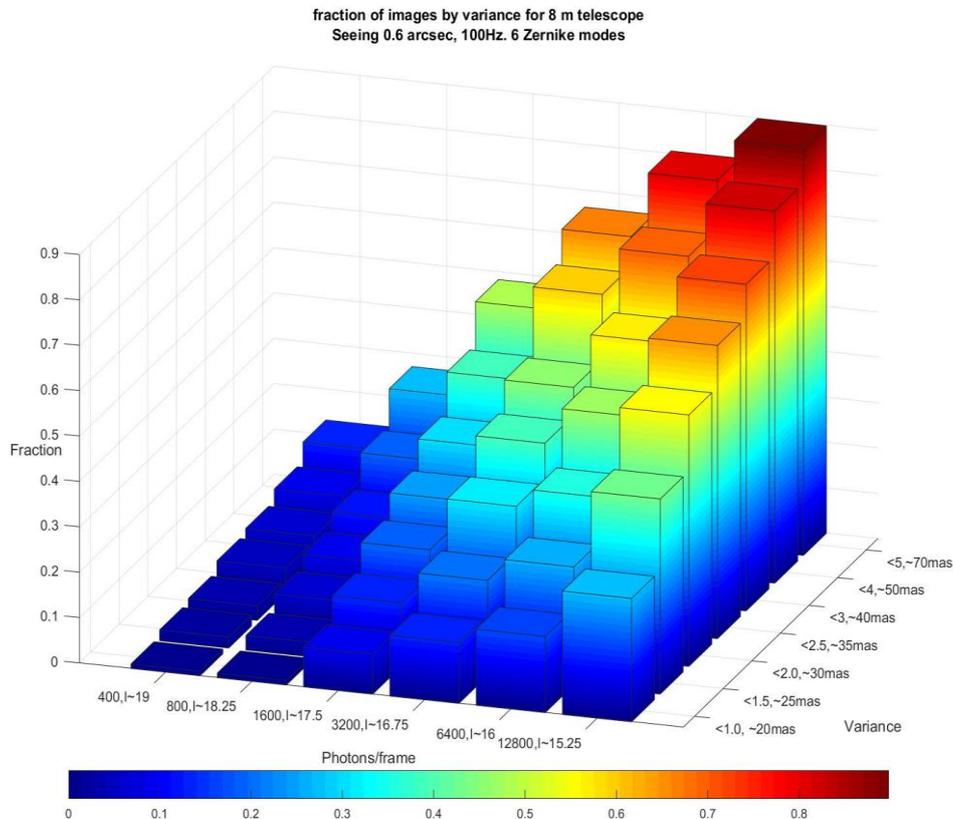

**Figure 10.** The fraction of images in different quality bins of variance compared with the photon levels per CWFS frame, together with an approximate estimate of the magnitude of star or compact galaxy that would produce such a photon rate on a 2.5 m telescope. The seeing was set to 0.6 arcseconds and 100 Hz wavefront sensor.

## 4    AOLI: Adaptive Optics Lucky Imager.

An instrument called AOLI, standing for Adaptive Optics Lucky Imager has been constructed for use on the 4.2 m William Herschel telescope on La Palma, in collaboration with the group at the Instituto de Astrofisica de Canarias (IAC) in Tenerife. This is presently undergoing commissioning. The principle of the design is that we keep the reference object on the optical axis of the telescope. A pickoff mirror is pierced with a small hole for the reference star but the remainder of the frame is folded onto the science detector array. The science detector array consists of four EMCCD devices with 1024 x 1024 pixels. These are run at high gain in photon counting mode and are precisely synchronised. These CCDs cannot be butted because of the package. However by using a shallow angle prism to separate the image in an intermediate focal plane beams are reimaged onto separated detectors to allow a combined field of view to be achieved.

The light for the curvature wavefront sensor is then reimaged and the light split to correspond to two separate equal distances on either side of the reimaged pupil. Images are then formed on another EMCCD which detects the defocused images formed by the telescope.

## 5      Adaptive Optics and Lucky Imaging on Even Larger Telescopes.

It is worth noting that there is no reason why these techniques should not be used on even larger telescopes in the infrared. As one goes to longer wavelengths the turbulent power in the wavefront is reduced approximately in proportion to the wavelength. It means that the same techniques can be used in the near infrared. Such an arrangement does call for high sensitivity K-band (2.2 μm) detectors running at a reasonably high frame rate though not as fast as used for the curvature wavefront sensors in the visible because the timescales over which the turbulence becomes decorrelated  are correspondingly longer in the IR.

## 6      Conclusions.

The use of curvature wavefront sensors is likely to revolutionise the detection of the phase errors caused by atmospheric turbulence on ground-based telescopes. By careful use of curvature wavefront sensors it is possible to work with dramatically fainter reference stars than are normally accessible with conventional Shack-Hartmann wavefront sensors. They will greatly increase the opportunities to take high resolution images in the visible on ground-based telescopes which exceed the sort of resolution that we have become so used to on the Hubble Space Telescope.

## 7      Acknowledgements

The author wishes to thank Marcos van Dam for enabling the author to use the Matlab software he developed with colleagues in New Zealand. The author also wishes to thank Olivier Guyon for many useful conversations over the years.

## 8      References

Baldwin et al., A&A, 480,589, 2008
Dekany, R. G. et al., SPIE, vol 3353, doi: 10.1117/12.321717
Guyon, O. et al., PASP, 120, 655, 2008.
Guyon,O., PASP,122, 49 2010.
Johnston, R., Ph D thesis, University of Canterbury, New Zealand, November 2000.
Noll, R. J.,1976, JOSA,66,207.
van Dam, M. A., Lane, R. G., Applied Optics , 41, 5497, 2002.